\newcommand*{\wn}{cm$^{-1}$}
\newcommand*{\AX}{A$^1\Pi-$X$^1\Sigma^+$}
\newcommand*{\dX}{d$^3\Delta-$X$^1\Sigma^+$}
\newcommand*{\eX}{e$^3\Sigma^--$X$^1\Sigma^+$}

\newcommand*{\ds}{d$^3\Delta$}
\newcommand*{\es}{e$^3\Sigma^-$}
\newcommand*{\as}{a$'^3\Sigma^+$}
\newcommand*{\Ds}{D$^1\Delta$}
\newcommand*{\Is}{I$^1\Sigma^-$}
\newcommand*{\As}{A$^1\Pi$}

\newcommand*{\Xs}{X$^1\Sigma^+$}

\newcommand*{\1}{$v=1$}
\newcommand*{\2}{$v=2$}
\newcommand*{\3}{$v=3$}
\newcommand*{\4}{$v=4$}
\newcommand*{\5}{$v=5$}
\newcommand*{\6}{$v=6$}
\newcommand*{\7}{$v=7$}
\newcommand*{\8}{$v=8$}

\newcommand*{\CO}{$^{12}$C$^{16}$O}
\documentclass[]{tMPH2e}
\usepackage{graphics}
\usepackage{dcolumn}
\usepackage[version=3]{mhchem}
\usepackage[english]{babel}
\usepackage{dcolumn}
\usepackage{graphicx}
\usepackage{amsmath}
\usepackage{tabularx}
\usepackage{bigstrut}
\usepackage{longtable}
\usepackage{rotating}
\usepackage{comment}
\usepackage{float}
\usepackage{verbatim}
\usepackage{color, soul}
\usepackage{threeparttable}
\usepackage{threeparttablex}

\begin{document}

\title{Spectroscopy and perturbation analysis of the CO \AX\ (2,0), (3,0) and (4,0) bands
}

\author{M. L. Niu$^{a}$${^\ast}$, E. J. Salumbides$^{a}$$^{b}$, A. N. Heays$^{c}$, N. de Oliveira$^{d}$\\R. W. Field$^{e}$, and W. Ubachs$^{a}$\thanks{$^\ast$Corresponding author. Email: m.niu@vu.nl
\vspace{6pt}}\\\vspace{6pt}  $^{a}${\em{Department of Physics and Astronomy, and LaserLaB, Vrije Universiteit, De Boelelaan 1081, 1081 HV Amsterdam, The Netherlands}};
$^{b}${\em{Department of Physics, University of San Carlos, Cebu City~6000, Philippines}};
$^{c}${\em{Leiden Observatory, Leiden University, PO Box 9513, 2300 RA Leiden, The Netherlands}};
$^{d}${\em{Synchrotron Soleil, Orme des Merisiers, St.~Aubin~BP~48, 91192, GIF~sur~Yvette~cedex, France}}; $^{e}${\em{Department of Chemistry, Massachusetts Institute of Technology, Cambridge, Massachusetts 02139, USA}}.\\\vspace{6pt}
}

\date{\today}
\maketitle

\begin{abstract}
\label{abstract}
The (2,0) (3,0) and (4,0) bands of the \AX\ system of \CO\ have been re-investigated by high-resolution vacuum ultraviolet absorption spectroscopy. A VUV Fourier-transform spectrometer, illuminated by synchrotron radiation, was applied to record a jet-cooled spectrum, a room temperature static gas spectrum and a high temperature (900 K) quasi-static gas spectrum, resulting in absolute accuracies of 0.01$-$0.02 \wn\ for the rotational line frequencies. Precise laser-based data were included in the analysis allowing for a highly accurate determination of band origins.
Rotational levels up to $J=52$ were observed. The data were used to perform an improved analysis of the perturbations in the \As, \2, \3, and \4\ levels by vibrational levels of the \Ds, \Is, \es, \ds, and \as\ states.

\begin{keywords}ultraviolet spectra; FT-spectroscopy; Doppler-free laser spectroscopy; perturbation analysis; carbon monoxide 
\end{keywords}\bigskip
\end{abstract}

\section{Introduction}

The spectroscopy of the carbon monoxide molecule remains of central interest to a variety of subfields in science. In particular the \AX\ system, investigated by a number of authors over decades~\cite{Birge1926, Herzberg1929, Simmons1969, Kepa1970, Lefloch1985, Field-thesis, Field1972a}, is often used as a probe for detecting CO. New and recent examples of its application are the proposal to search for a varying proton-electron mass ratio on cosmological time scales~\cite{Salumbides2012}, and to probe the local cosmic microwave background temperature as a function of redshift~\cite{Noterdaeme2011}.
For these applications the analysis of the \AX\ system is warranted at the highest accuracy. At the same time the \As\ state of CO is known as
a celebrated example of perturbations, which makes its study interesting from a pure molecular physics perspective. A first comprehensive perturbation analysis was performed by Field~\emph{et al.}~\cite{Field-thesis, Field1972a}.

After having performed an improved perturbation analysis for the \AX\ $(0,0)$ and $(1,0)$ bands~\cite{Niu2013} we here extend the updated
perturbation analysis to the \AX\ $(2,0)$, $(3,0)$ and $(4,0)$ bands by using the high resolution vacuum ultraviolet (VUV) Fourier-transform (FT) spectroscopy setup at the DESIRS beamline at the Soleil synchrotron. For the purpose of achieving an absolute wavelength calibration of the rotational lines, a subset of lines was first probed with laser-based Doppler-free two-photon spectroscopy~\cite{Niu2015}. For the present study the VUV-FT instrument was used in three different modes of operation: gas-jet spectroscopy, room-temperature static gas absorption, and absorption at 900 K, for which a special setup was designed~\cite{Niu2015a}.
The combination of these measurements allows for a highly accurate analysis of the spectrum of the three bands, probing rotational states as high as $J=52$, providing information on perturbing states interacting at the high rotational quantum numbers. The aim of using these different configurations will be discussed in the next section.

\section{Experimental details}
The vacuum ultraviolet (VUV) Fourier-transform (FT) spectrometer at the DESIRS beamline of the SOLEIL synchrotron is a unique tool for recording high-resolution absorption spectra in the range $4-30$ eV~\cite{Deoliveira2009,Deoliveira2011}.
For the present investigation, the instrument was used in three configurations, each being a compromise between obtaining narrow linewidths and high wavelength accuracy, or probing as many rotational levels as possible.
The free-jet configuration is used to record the narrowest transitions. First, in the jet-expansion, the perpendicular directionality of the molecular beam gives rise to a much reduced Doppler width, yielding an observed width of 0.09 \wn\ in combination with the instrument settings of the FT-spectrometer.
Under the jet conditions the rotational temperature is reduced to 12 K and only rotational levels $J=0-5$  are probed at this high resolution. 

\begin{figure}
\begin{center}
\resizebox{0.8\textwidth}{!}{\includegraphics{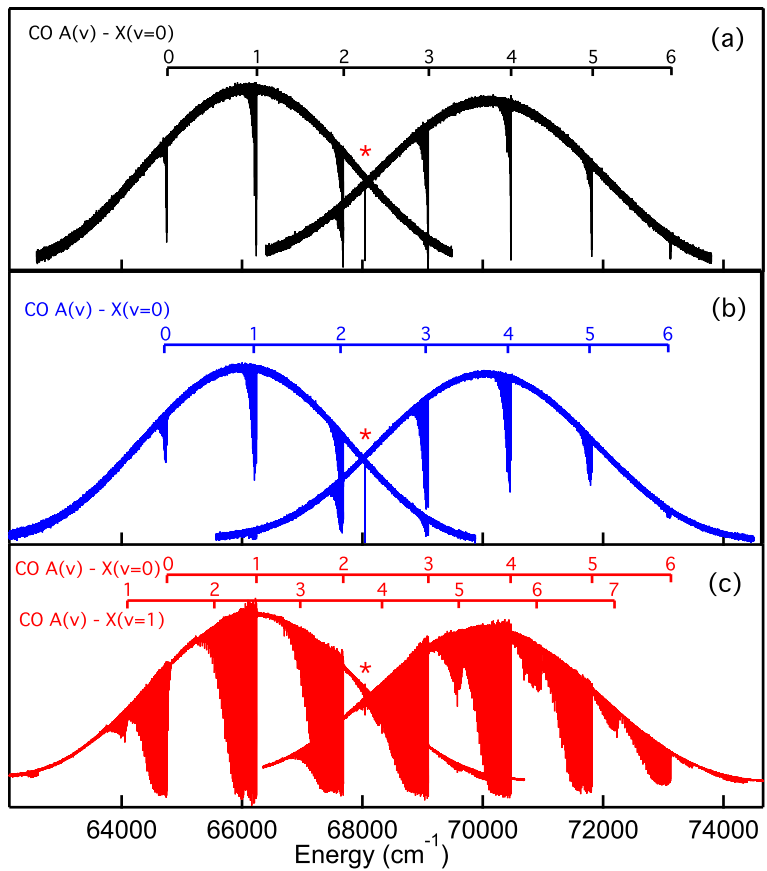}}
\caption{Overview spectra of the CO \AX\ system including $(2,0)$, $(3,0)$ and $(4,0)$ bands recorded with the vacuum ultraviolet Fourier-transform spectrometer at the SOLEIL synchrotron under three different experimental conditions; (a) free molecular jet expansion; (b) room temperature quasi-static gas cell; (c)  a free-flowing gas cell heated to 900 K. The asterisk (*) indicates the Xe atomic resonance line used in for calibration of the FTS instrument.}
\label{fig:FTSspec}
\end{center}
\end{figure}

Second, FT-spectra are recorded under quasi-static room-temperature conditions with the use of a windowless cell. In this configuration the linewidth obtained is 0.16 \wn~\cite{Niu2013}, while rotational lines up to $J\sim20$ are observed.
In a third configuration, a windowless gas cell was heated up to 900 K, in order to record the highest rotational quantum states. The linewidth under these conditions, at full width of half maximum (FWHM), was 0.39 \wn~\cite{Niu2015a}. The latter spectra were recorded at relatively high column densities, which is around a factor of 100 higher compared to the sample used with the unheated cell. It is used to probe rotational states with the highest $J$ quantum number, in which case the low-$J$ transitions are saturated. In all the FT-experiments, CO gas was used at a purity of 99.997\% from Air Liquide, presumably composed of the regular terrestrial $^{12}$C/$^{13}$C and $^{16}$O/$^{17}$O/$^{18}$O isotopic abundances.

\begin{figure*}
\begin{center}
\resizebox{1\textwidth}{!}{\includegraphics{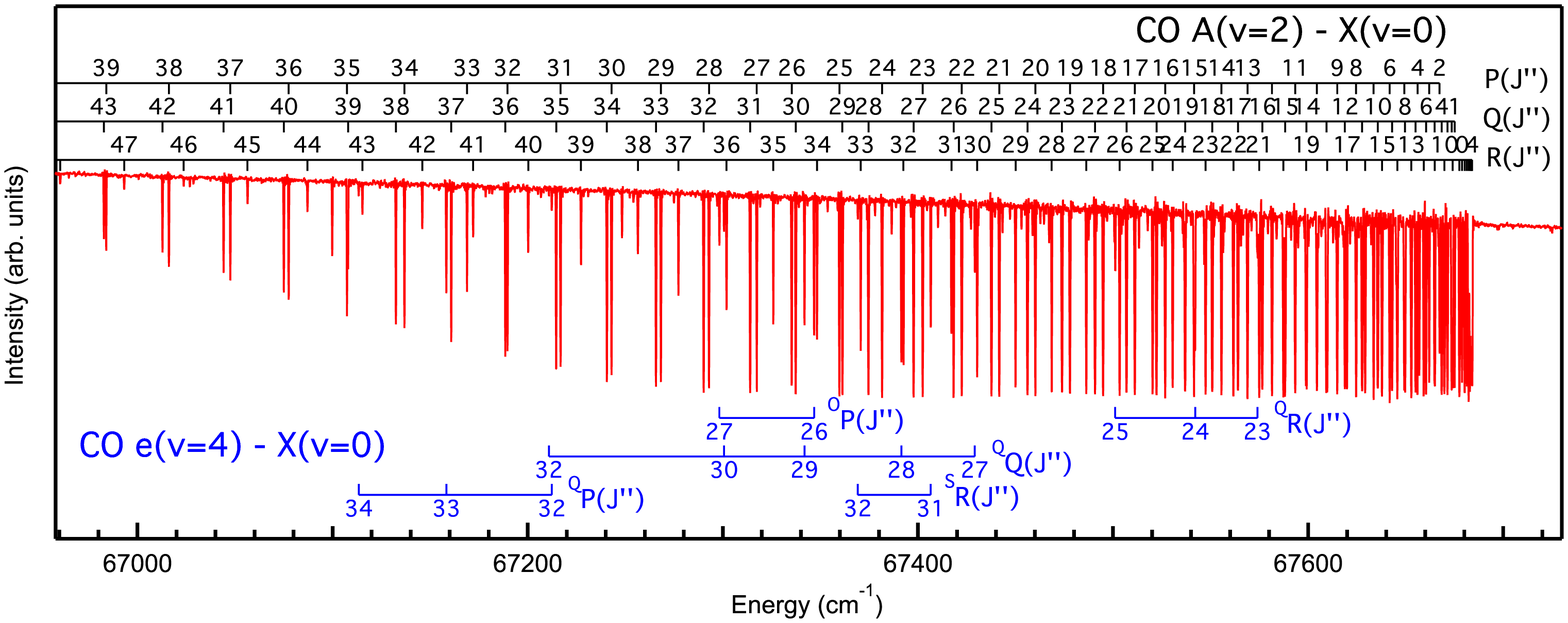}}
\caption{A zoom-in spectrum recorded with a windowless gas cell heated to 900 K showing absorption lines in the \AX\ (2,0) band as well as perturber lines belonging to the \es\ $(4,0)$ band.}
\label{fig:Hotspec}
\end{center}
\end{figure*}

Figure~\ref{fig:FTSspec} shows characteristic overview VUV-FT spectra recorded under the three measurement conditions, covering the range of the \AX $(v',0)$ bands for $v'=0$ to $v'=6$.
The high temperature spectrum also shows hot bands, i.e. \AX\ $(v'=1-7,1)$, originating from the \Xs, $v"=1$ state.
The $(2,1)$, $(3,1)$, and $(4,1)$ hot bands are expected to be weak due to small Franck-Condon factors~\cite{Beegle1999}.
A zoom-in part of the high temperature spectrum is presented in Fig.~\ref{fig:Hotspec}, showing transitions in the \AX\ (2,0) band and some perturber lines belonging to the \eX\ $(4,0)$ band. The unlabeled weak lines belong to excitation of other perturber states, mentioned below.

In order to obtain the most accurate transition frequencies,  different settings on the FT-instrument are used in combination with the various measurement configurations. The free-jet and room temperature spectra are both recorded by taking 1978 kilo-samples of data over the 0 to 40 mm optical path difference within the interferometer, yielding an instrumental resolution of 0.075 \wn, corresponding to the ultimate resolving power of the instrument. For the hot cell spectra, because of the increased Doppler broadening of about 0.28 \wn, constraints on the instrumental resolution are relaxed to 0.27 \wn by taking 1024 kilo-samples of interferometric data to save recording time, which permits more averaging, thus increasing the signal-to-noise ratio. The absolute calibration for all the FT spectra is obtained from on-line recording of a xenon line at 68\,045.156\,(3) \wn~\cite{Saloman2004}.

\section{Results}

Collectively, more than 450 absorption lines are observed in the region from 66\,400 to 70\,500 \wn, including rotational levels up to $J=52$ for the main \AX\ bands, as well as a large number of transitions belonging to perturber states.
In Table~\ref{Astatetransition}, transition frequencies in the CO \AX\ $(2,0)$, $(3,0)$ and $(4,0)$ bands are listed. The absorption lines associated with excitation of the perturber states are presented in Table~\ref{perturbingstatetransition}.
In these tables, the subscripts $e$ and $f$  denote the electronic symmetry of the upper state. The superscripts Q, S, R, O and P in Table~\ref{perturbingstatetransition}, indicate the change in total angular momentum excluding spin for transitions to perturber states~\cite{Morton1994}.
The uncertainties of transition frequencies are 0.02 \wn\ for most of the transitions, except the weak or blended ones.
To verify the accuracy of the FT data, we compare combination differences between P$(J)$ and R$(J-2)$ transitions in the FT data with the very accurate far-infrared data~\cite{Varberg1992}. The comparison yields good agreement with a standard deviation of 0.015 \wn, smaller than the estimated FT uncertainty.

\begin{table*}[t]
\begin{center}
\begin{scriptsize}
\caption{Transition frequencies (in vacuum \wn) in the $^{12}$C$^{16}$O $A^1\Pi$ - $X^1\Sigma^+$ $(2,0)$, $(3,0)$, and $(4,0)$ bands obtained in the present VUV-FT experiment and absolutely calibrated with respect to the laser data~\cite{Niu2015}. $J''$ is the rotational quantum number in the ground state. The subscripts $e$ and $f$ indicate the electronic symmetry of the upper state. The superscripts \textit{b} and \textit{w} indicate blended and weak transitions, respectively.}
\label{Astatetransition}
\begin{tabular}
{c@{\hspace{5pt}}l@{\hspace{5pt}}l@{\hspace{5pt}}l@{\hspace{5pt}}l@{\hspace{5pt}}l@{\hspace{5pt}}l@{\hspace{5pt}}l@{\hspace{5pt}}l@{\hspace{5pt}}l}
\hline
\hline\\[-2ex]
 & \multicolumn{3}{c}{(2,0)} & \multicolumn{3}{c}{(3,0)} & \multicolumn{3}{c}{(4,0)}  \\
\hline\\[-2ex]
$J''$	&	R$_e(J'')$	&	Q$_f(J'')$	&	P$_e(J'')$	&	R$_e(J'')$	&	Q$_f(J'')$	&	P$_e(J'')$	&	R$_e(J'')$	&	Q$_f(J'')$	&	P$_e(J'')$	\\
0	&	67678.89 	&		&		&	69091.63 	&		&		&	70469.93 	&		&		\\
1	&	67681.28 	&	67675.05 	&		&	69093.91 	&	69087.79 	&		&	70472.08 	&	70466.11 	&		\\
2	&	67682.94 	&	67673.59 	&	67667.35 	&	69095.43 	&	69086.24 	&	69080.10 	&	70473.41 	&	70464.44 	&	70458.45 	\\
3	&	67683.86 	&	67671.40 	&	67662.06 	&	69096.18$^b$ 	&	69083.91 	&	69074.71 	&	70473.86 	&	70461.93 	&	70452.85 	\\
4	&	67684.06 	&	67668.48 	&	67656.02 	&	69096.13$^b$ 	&	69080.80 	&	69068.53 	&	70473.50$^b$ 	&	70458.57 	&	70446.47 	\\
5	&	67683.51 	&	67664.83$^b$ 	&	67649.25 	&	69095.32$^b$ 	&	69076.91 	&	69061.58 	&	70472.35 	&	70454.36$^b$ 	&	70439.26 	\\
6	&	67682.25 	&	67660.41 	&	67641.77 	&	69093.72 	&	69072.25 	&	69053.85 	&	70470.37 	&	70449.38 	&	70431.23 	\\
7	&	67680.25 	&	67655.09 	&	67633.55 	&	69091.35 	&	69066.81 	&	69045.35 	&	70467.58 	&	70443.55 	&	70422.38 	\\
8	&	67677.53 	&	67649.64 	&	67624.60 	&	69088.20 	&	69060.61$^b$ 	&	69036.07 	&	70463.96 	&	70436.90 	&	70412.72 	\\
9	&	67674.06 	&	67643.01 	&	67614.92 	&	69084.27 	&	69053.61 	&	69026.02 	&	70459.53 	&	70429.43 	&	70402.26$^b$ 	\\
10	&	67669.87 	&	67635.69 	&	67604.51 	&	69079.56 	&	69045.84 	&	69015.19 	&	70454.30$^b$ 	&	70421.15 	&	70390.95 	\\
11	&	67664.92$^b$ 	&	67627.65 	&	67593.38 	&	69074.07 	&	69037.29 	&	69003.58 	&	70448.21 	&	70412.06 	&	70378.84 	\\
12	&	67659.27 	&	67618.88 	&	67581.50 	&	69067.80 	&	69027.97 	&	68991.20 	&	70441.31 	&	70402.18$^b$ 	&	70365.92 	\\
13	&	67652.87 	&	67609.38 	&	67568.90 	&	69060.76$^b$ 	&	69017.87 	&	68978.04 	&	70433.59 	&	70391.41 	&	70352.18 	\\
14	&	67645.74 	&	67599.14 	&	67555.57 	&	69052.92 	&	69006.99 	&	68964.12 	&	70425.05 	&	70379.85 	&	70337.61 	\\
15	&	67637.86 	&	67588.18 	&	67541.51 	&	69044.32 	&	68995.33 	&	68949.40 	&	70415.67 	&	70367.48 	&	70322.23 	\\
16	&	67629.24 	&	67576.49 	&	67526.73 	&	69034.92 	&	68982.90 	&	68933.92 	&	70405.47 	&	70354.27 	&	70306.04 	\\
17	&	67619.89 	&	67564.05 	&	67511.19 	&	69024.77 	&	68969.67 	&	68917.65 	&	70394.44 	&	70340.24 	&	70289.00 	\\
18	&	67609.78 	&	67550.87 	&	67494.92 	&	69013.78 	&	68955.70 	&	68900.60 	&	70382.59 	&	70325.39 	&	70271.16 	\\
19	&	67598.95$^b$ 	&	67536.94 	&	67477.92 	&	69002.03 	&	68940.89 	&	68882.80 	&	70369.88 	&	70309.72 	&	70252.48 	\\
20	&	67587.33 	&	67522.30 	&	67460.19 	&	68989.48 	&	68925.33 	&	68864.21 	&	70356.37 	&	70293.21 	&	70233.00 	\\
21	&	67574.92 	&	67506.89 	&	67441.70 	&	68976.22 	&	68908.97 	&	68844.79 	&	70342.03 	&	70275.86 	&	70212.65 	\\
22	&	67561.70 	&	67490.78 	&	67422.46 	&	68962.08 	&	68891.89 	&	68824.64 	&	70326.84 	&	70257.68 	&	70191.52 	\\
23	&	67547.46 	&	67473.82 	&	67402.45 	&	68947.17 	&	68873.96 	&	68803.75 	&	70310.78 	&	70239.24 	&	70169.54 	\\
24	&	67530.58 	&	67456.15 	&	67381.62 	&	68931.46 	&	68855.22 	&	68781.99 	&	70293.93 	&	70218.90 	&	70146.73 	\\
25	&	67520.28 	&	67437.66 	&	67359.77 	&	68915.02 	&	68835.70 	&	68759.48 	&	70276.21 	&	70198.21 	&	70123.10 	\\
26	&	67503.39 	&	67418.29 	&	67335.29 	&	68897.69 	&	68815.44 	&	68736.18 	&	70257.68 	&	70176.71 	&	70098.63 	\\
27	&	67486.31 	&	67397.77 	&	67317.41 	&	68879.56 	&	68794.20 	&	68712.16 	&	70238.31$^b$ 	&	70154.37 	&	70073.34 	\\
28	&	67468.58 	&	67374.63 	&	67292.93 	&	68860.55 	&	68770.99 	&	68687.22 	&	70218.03 	&	70131.19 	&	70047.21 	\\
29	&	67450.03 	&	67361.31 	&	67268.28 	&	68840.30 	&	68750.33 	&	68661.52 	&	70196.95 	&	70107.17 	&	70020.24 	\\
30	&	67430.37 	&	67337.38 	&	67242.97 	&	68822.27 	&	68726.69 	&	68634.95 	&	70175.00 	&	70082.31 	&	69992.43 	\\
31	&	67417.14 	&	67314.00 	&	67216.87 	&	68800.01 	&	68702.36 	&	68607.14 	&	70152.17 	&	70056.60 	&	69963.79 	\\
32	&	67392.61 	&	67290.19 	&	67189.66 	&	68777.81 	&	68677.19 	&	68581.56 	&	70128.35 	&	70030.04 	&	69934.29 	\\
33	&	67370.68 	&	67265.74 	&	67168.90 	&	68754.92 	&	68651.02 	&	68551.76 	&	70106.17 	&	70002.63 	&	69903.92 	\\
34	&	67348.30 	&	67240.58 	&	67136.83 	&	68731.22 	&	68625.59 	&	68522.08 	&	70079.26 	&	69974.35 	&	69872.57 	\\
35	&	67325.89 	&	67214.61 	&	67107.38 	&	68706.74 	&	68597.80 	&	68491.62 	&	70052.93 	&	69945.17 	&	69842.87 	\\
36	&	67301.93 	&	67188.51 	&	67077.50 	&	68681.45 	&	68569.49 	&	68460.41 	&	70025.84 	&	69914.90 	&	69808.45 	\\
37	&	67277.26 	&	67160.83 	&	67047.57 	&	68655.36 	&	68540.43 	&	68428.42 	&	69997.89 	&	69888.36 	&	69774.62 	\\
38	&	67256.49 	&	67132.43 	&	67016.12 	&	68628.46 	&	68510.57 	&	68395.65 	&	69969.03 	&	69853.81 	&	69740.03 	\\
39	&	67227.31 	&	67107.93 	&	66983.97 	&	68600.75 	&	68479.90 	&	68362.08 	&	69938.72 	&	69821.03 	&	69704.61 	\\
40	&	67200.28 	&	67075.00 	&	66955.71 	&	68572.22 	&	68448.23 	&	68327.76 	&	69909.59 	&	69787.59 	&	69668.29 	\\
41	&	67172.06 	&	67044.14 	&	66919.10 	&	68542.87 	&	68416.52 	&	68292.54 	&	69878.01 	&	69753.32 	&	69630.43$^b$ 	\\
42	&	67145.95 	&	67012.83 	&	66884.61 	&	68512.68 	&	68383.32 	&	68256.55$^b$ 	&	69845.70 	&	69718.20 	&	69593.95 	\\
43	&	67115.28 	&	66982.80 	&	66848.98 	&	68481.65 	&	68349.41 	&	68219.80 	&	69812.54 	&	69682.03 	&	69554.93 	\\
44	&	67087.12 	&	66949.46 	&	66815.45 	&	68449.64 	&	68314.71 	&	68182.17 	&	69778.25 	&	69646.17$^b$ 	&	69515.21 	\\
45	&	67056.40 	&	66915.88 	&	66777.38 	&	68415.42 	&	68279.13 	&	68143.74 	&	69747.15 	&	69607.90$^b$ 	&	69474.61 	\\
46	&	67023.71 	&	66881.55 	&	66741.82 	&	68384.81 	&	68242.75 	&	68104.34$^b$ 	&	69708.67 	&	69569.50 	&	69432.95$^w$ 	\\
47	&	66993.21 	&	66703.75 	&	66703.75 	&	68350.09 	&	68205.49 	&	68062.70$^w$ 	&		&	69530.04 	&	69394.53 	\\
48	&	66960.40 	&	66811.29 	&		&		&	68167.17$^b$ 	&	68024.82$^w$ 	&		&	69489.51 	&	69348.60$^b$ 	\\
49	&	66926.80 	&	66774.68 	&	66625.81$^w$ 	&		&	68126.16$^w$ 	&		&	69596.13 	&	69449.25 	&	69304.46$^w$ 	\\
50	&	66892.44 	&	66737.33 	&	 	&		&		&		&		&	69407.01 	&		\\
51	&	66857.29$^w$ 	&	66699.10 	&	 	&		&		&		&		&	69364.03$^w$ 	&		\\
52	&	66821.38$^w$ 	&		&		&		&		&		&		&		&		\\
\hline
\hline
\end{tabular}
\end{scriptsize}
\end{center}
\end{table*}

\renewcommand\arraystretch{1}
\begin{table*}[htp]
\begin{center}
\begin{small}
\caption{Transition frequencies (in vacuum \wn) for excitation of the various perturber states obtained in the present VUV-FT measurements, recalibrated from the laser data. The quantum number in brackets $J''$ represents the total angular momentum of the ground state. The left-superscripts Q, S, R, O and P denote the total angular momentum excluding spin of the perturber states, according to the notation in Ref.~\cite{Morton1994}. The subscripts $e$ and $f$ indicate the electronic symmetry of the upper state. The superscripts \textit{b} and \textit{w} indicate blended and weak transitions, respectively.}
\label{perturbingstatetransition}
\begin{tabular}
{l@{\hspace{20pt}}l@{\hspace{15pt}}l@{\hspace{20pt}}l@{\hspace{15pt}}l@{\hspace{20pt}}l}
\hline
\hline\\[-2ex]
\multicolumn{2}{c}{$d^3\Delta - X^1\Sigma^+$ (7,0)}			&	\multicolumn{2}{l}{$e^3\Sigma^- - X^1\Sigma^+$ (4,0)}		&	\multicolumn{2}{l}{$I^1\Sigma^- - X^1\Sigma^+$ (3,0)}			\\
$^R$Q$_f$(42)							&	67028.66$^w$	&	$^O$P$_e$(26)	&	67346.78		&	$^Q$Q$_f$(6)	&	67666.78$^w$	\\
$^R$Q$_f$(43)							&	66964.86	&	$^O$P$_e$(27)	&	67298.26		&	$^Q$Q$_f$(7)	&	67656.98	\\
$^P$P$_e$(40)							&	66947.61	&	$^Q$R$_e$(23)	&	67574.05$^w$		&	$^Q$Q$_f$(8)	&	67645.13$^b$	\\
$^R$R$_e$(38)							&	67248.35	&	$^Q$R$_e$(24)	&	67542.09		&	$^Q$Q$_f$(9)	&	67632.32$^w$	\\
$^Q$Q$_f$(39)							&	67099.82	&	$^Q$R$_e$(25)	&	67501.14		&			&\\
								&			&	$^Q$Q$_f$(27)	&	67429.16		&	\multicolumn{2}{l}{$I^1\Sigma^- - X^1\Sigma^+$ (6,0)}			\\
\multicolumn{2}{c}{$d^3\Delta - X^1\Sigma^+$ (8,0)}					&	$^Q$Q$_f$(28)	&	67391.50		&	$^Q$Q$_f$(23)	&	70237.93$^b$					\\
$^S$R$_e$(25)							&	68912.94$^w$	&	$^Q$Q$_f$(29)	&	67341.85		&			&\\
$^R$Q$_f$(26)							&	68813.41	&	$^Q$Q$_f$(30)	&	67300.60		&	\multicolumn{2}{l}{$a'^3\Sigma^+ - X^1\Sigma^+$ (13,0)}			\\
								&			&	$^Q$Q$_f$(32)	&	67210.86$^w$		&	$^P$P$_e$(31)	&	68624.61					\\
\multicolumn{2}{c}{$d^3\Delta - X^1\Sigma^+$ (10,0)}					&	$^Q$P$_e$(32)	&	67212.33		&	$^P$P$_e$(32)	&	68568.47					\\
$^Q$Q$_f$(45)							&	69607.35$^b$	&	$^Q$P$_e$(33)	&	67158.32		&	$^R$R$_e$(29)	&	68857.80					\\
								&			&	$^Q$P$_e$(34)	&	67113.40		&	$^R$R$_e$(30)	&	68809.19					\\
\multicolumn{2}{c}{$D^1\Delta - X^1\Sigma^+$ (3,0)}					&	$^S$R$_e$(31)	&	67406.56		&	$^R$Q$_f$(33)	&	68664.38					\\
$^P$P$_e$(48)							&	66666.80	&	$^S$R$_e$(32)	&	67369.20		&	$^R$Q$_f$(34)	&	68610.50					\\
$^R$R$_e$(46)							&	67026.86	&			&				&		&\\
								&			&\multicolumn{2}{l}{$e^3\Sigma^- - X^1\Sigma^+$ (7,0)} 	&	\multicolumn{2}{l}{$a'^3\Sigma^+ - X^1\Sigma^+$ (15,0)}			\\
								&			&	$^O$P$_e$(35)	&	69836.03		&	$^P$P$_e$(47)	&	69386.27$^w$					\\
								&			&	$^Q$R$_e$(33)	&	70099.33		&	$^R$R$_e$(45)	&	69738.97					\\
								&			&	$^Q$Q$_f$(36)	&	69941.07$^w$		&			&	\\
								&			&	$^Q$Q$_f$(37)	&	69879.81		&			&		\\
								&			&	$^S$R$_e$(39)	&	69948.44$^w$ &&\\
\hline                                                                                          			
\hline
\end{tabular}
\end{small}
\end{center}
\end{table*}

\begin{figure}
\begin{center}
\resizebox{0.8\textwidth}{!}{\includegraphics{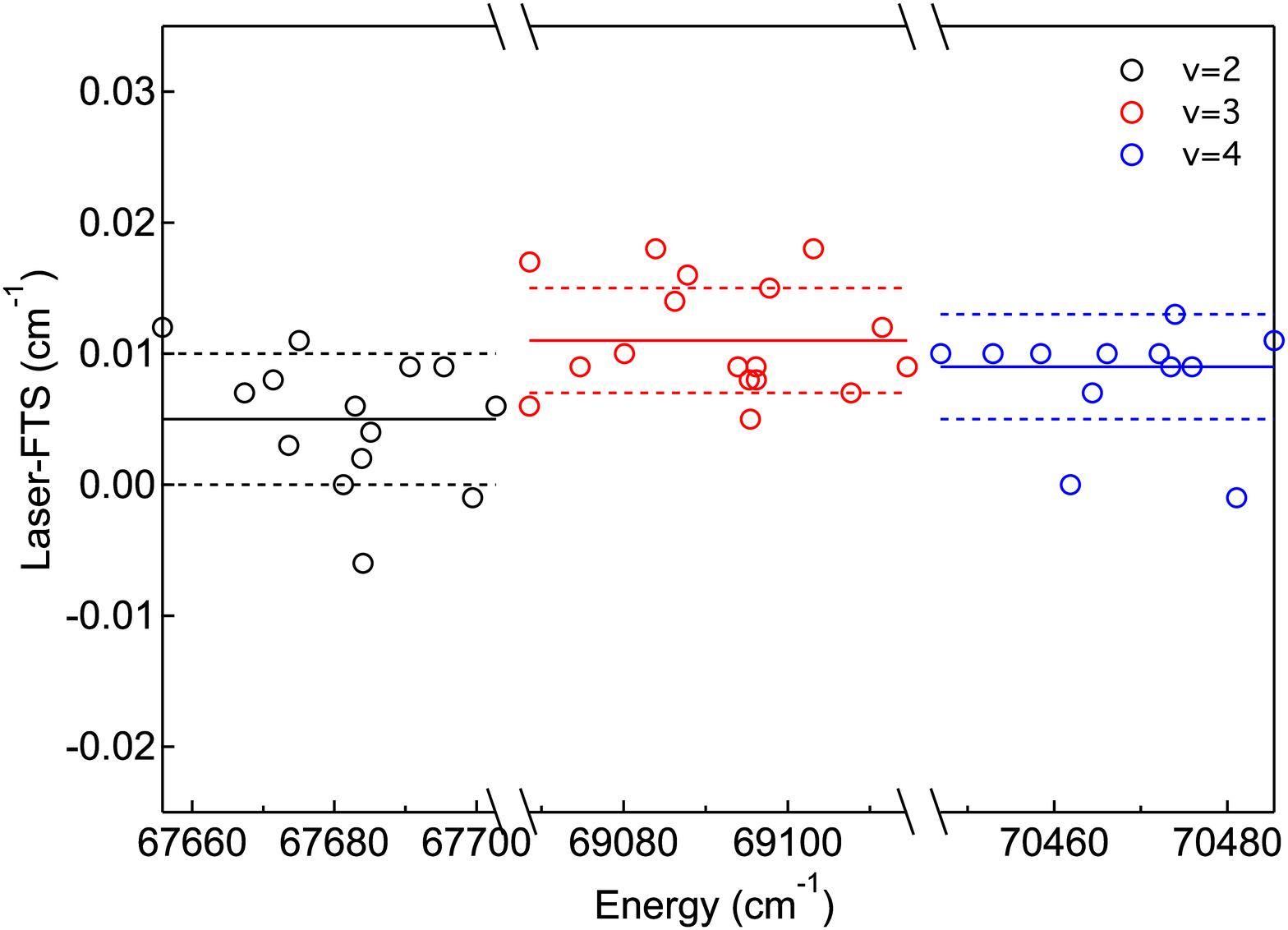}}
\caption{The difference between \As\ level energies derived from the laser and FTS datasets for low $J$ levels.
The energy differences are applied as corrections to FTS-derived level energies and transition frequencies reported here.}
\label{laser-FTS}
\end{center}
\end{figure}

In view of parity selection rules, the measured transition frequencies in the present one-photon absorption experiment cannot be compared directly with values obtained in the two-photon laser experiment for the same bands~\cite{Niu2015}.
However, based on accurately known ground-state level energies [18] and the derived excited state $\Lambda-$doublet splittings, a verification of absolute level energies derived from the VUV-FT experiment can be compared with the more accurate data from the laser experiment, yielding the differences shown in Fig.~\ref{laser-FTS}.
The average offsets between the two data sets are different for different bands, on the order of 0.01 \wn. These small discrepancies are attributed to an offset in the FT data that is well within its estimated uncertainty. The standard deviation of $\sim0.005$ \wn\ demonstrates that the relative uncertainty of the FT data is much smaller than the estimated absolute uncertainty.
The energy offsets with respect to the more accurate laser spectroscopic data were used to correct the level energies of the \As\ $v=2,3,4$ levels by 0.005, 0.011 and 0.009 \wn, respectively.
The corrected level energies are listed in Table~\ref{Astatelevelenergies}, where the values derived from laser data are used for $J=1-6$.
The level energies of perturber states are also corrected by the calibration shift and shown in Table~\ref{perturbingstatelevelenergies}.
In a similar way the corrections for the level energies, have also been applied to the transition energies listed in Tables~\ref{Astatetransition}~and~\ref{perturbingstatetransition}.

\renewcommand\arraystretch{1}
\begin{table*}[t]
\begin{center}
\begin{scriptsize}
\caption{Level energies (in vacuum \wn) of \As\ $v=2,3,4$ states. The energies indicated with * are the level energies from the laser data~\cite{Niu2013}. The FT data include energy corrections deduced from the comparison with laser data.}
\label{Astatelevelenergies}
\begin{tabular}
{c@{\hspace{15pt}}c@{\hspace{5pt}}c@{\hspace{15pt}}c@{\hspace{5pt}}c@{\hspace{15pt}}c@{\hspace{5pt}}c}
\hline
\hline\\[-2ex]
 & \multicolumn{2}{c}{$v=2$} & \multicolumn{2}{c}{$v=3$} & \multicolumn{2}{c}{$v=4$}  \\
\hline\\[-2ex]
$J'$	&	$e$	&	$f$	&	$e$	&	$f$	&	$e$	&	$f$	\\
1	&	67678.895\,* 	&	67678.895\,* 	&	69091.640\,* 	&	69091.639\,* 	&	70469.933\,* 	&	70469.951\,* 	\\
2	&	67685.127\,* 	&	67685.123\,* 	&	69097.775\,* 	&	69097.775\,* 	&	70475.924\,* 	&	70475.971\,* 	\\
3	&	67694.474\,* 	&	67694.467\,* 	&	69106.979\,* 	&	69106.979\,* 	&	70484.923\,* 	&	70484.995\,* 	\\
4	&	67706.932\,* 	&	67706.922\,* 	&	69119.251\,* 	&	69119.249\,* 	&	70496.934\,* 	&	70497.023\,* 	\\
5	&	67722.506\,* 	&	67722.486\,* 	&	69134.588\,* 	&	69134.585\,* 	&	70511.96 	&	70512.03 	\\
6	&	67741.192\,* 	&	67741.14 	&	69152.994\,* 	&	69152.988\,* 	&	70530.02 	&	70530.12 	\\
7	&	67762.99 	&	67762.73 	&	69174.46 	&	69174.46 	&	70551.11 	&	70551.19 	\\
8	&	67787.90 	&	67788.03 	&	69198.99 	&	69199.00 	&	70575.23 	&	70575.29 	\\
9	&	67815.92 	&	67815.99 	&	69226.59 	&	69226.59 	&	70602.35 	&	70602.41 	\\
10	&	67847.04 	&	67847.09 	&	69257.25 	&	69257.24 	&	70632.51 	&	70632.56 	\\
11	&	67881.27 	&	67881.31 	&	69290.97 	&	69290.96 	&	70665.69 	&	70665.72 	\\
12	&	67918.59 	&	67918.65 	&	69327.74 	&	69327.74 	&	70701.88 	&	70701.94 	\\
13	&	67959.04 	&	67959.07 	&	69367.57 	&	69367.57 	&	70741.07 	&	70741.11 	\\
14	&	68002.56 	&	68002.61 	&	69410.46 	&	69410.45 	&	70783.29 	&	70783.32 	\\
15	&	68049.20 	&	68049.24 	&	69456.39 	&	69456.39 	&	70828.51 	&	70828.53 	\\
16	&	68098.91 	&	68098.96 	&	69505.37 	&	69505.37 	&	70876.72 	&	70876.75 	\\
17	&	68151.71 	&	68151.77 	&	69557.39 	&	69557.40 	&	70927.95 	&	70927.96 	\\
18	&	68207.60 	&	68207.66 	&	69612.48 	&	69612.49 	&	70982.16 	&	70982.18 	\\
19	&	68266.57 	&	68266.62 	&	69670.58 	&	69670.57 	&	71039.38 	&	71039.40 	\\
20	&	68328.62 	&	68328.68 	&	69731.70 	&	69731.71 	&	71099.56 	&	71099.59 	\\
21	&	68393.70 	&	68393.80 	&	69795.87 	&	69795.87 	&	71162.75 	&	71162.76 	\\
22	&	68461.82 	&	68462.01 	&	69863.12 	&	69863.12 	&	71228.92 	&	71228.92 	\\
23	&	68532.93 	&	68533.19 	&	69933.31 	&	69933.33 	&	71298.06 	&	71298.61 	\\
24	&	68606.83 	&	68607.46 	&	70006.54 	&	70006.54 	&	71370.16 	&	71370.21 	\\
25	&	68681.89 	&	68684.72 	&	70082.78 	&	70082.76 	&	71445.24 	&	71445.27 	\\
26	&	68767.34 	&	68764.89 	&	70162.09 	&	70162.04 	&	71523.27 	&	71523.31 	\\
27	&	68849.99 	&	68847.71 	&	70244.28 	&	70244.14 	&	71604.27 	&	71604.30 	\\
28	&	68936.25 	&	68931.69 	&	70329.49 	&	70328.05 	&	71688.23 	&	71688.25 	\\
29	&	69025.64 	&	69029.28 	&	70417.61 	&	70418.30 	&	71775.09 	&	71775.14 	\\
30	&	69118.00 	&	69120.04 	&	70508.27 	&	70509.35 	&	71864.92 	&	71864.97 	\\
31	&	69213.03 	&	69215.13 	&	70604.93 	&	70603.49 	&	71957.66 	&	71957.73 	\\
32	&	69318.27 	&	69313.57 	&	70701.14 	&	70700.56 	&	72053.30 	&	72053.41 	\\
33	&	69415.98 	&	69415.12 	&	70801.21 	&	70800.40 	&	72151.72 	&	72152.01 	\\
34	&	69520.06 	&	69519.73 	&	70904.30 	&	70904.74 	&	72255.55 	&	72253.50 	\\
35	&	69627.46 	&	69627.29 	&	71010.37 	&	71010.48 	&	72358.41 	&	72357.85 	\\
36	&	69738.57 	&	69738.47 	&	71119.41 	&	71119.45 	&	72465.61 	&	72464.86 	\\
37	&	69851.89 	&	69851.82 	&	71231.41 	&	71231.42 	&	72575.80 	&	72579.35 	\\
38	&	69968.25 	&	69968.19 	&	71346.35 	&	71346.34 	&	72688.88 	&	72689.57 	\\
39	&	70092.23 	&	70092.20 	&	71464.24 	&	71464.17 	&	72804.79 	&	72805.30 	\\
40	&	70211.58 	&	70211.51 	&	71585.02 	&	71584.74 	&	72922.94 	&	72924.10 	\\
41	&	70336.78 	&	70336.61 	&	71708.72 	&	71708.99 	&	73046.10 	&	73045.79 	\\
42	&	70464.53 	&	70464.99 	&	71835.35 	&	71835.47 	&	73170.48 	&	73170.35 	\\
43	&	70598.10 	&	70598.35 	&	71964.83 	&	71964.96 	&	73297.86 	&	73297.58 	\\
44	&	70730.83 	&	70732.11 	&	72097.20 	&	72097.37 	&	73428.08 	&	73428.82 	\\
45	&	70869.77 	&	70869.34 	&	72232.29 	&	72232.59 	&	73560.90 	&	73561.36 	\\
46	&	71009.87 	&	71009.51 	&	72368.85 	&	72370.71 	&	73700.64 	&	73697.45 	\\
47	&	71151.67 	&	71009.89 	&	72512.79 	&	72511.62 	&	73836.61 	&	73836.18 	\\
48	&	71299.35 	&	71299.30 	&	72656.23 	&	72655.17 	&	73978.00 	&	73977.51 	\\
49	&	71448.40 	&	71448.22 	&	69863.12 	&	72799.70 	&		&	74122.79 	\\
50	&	71600.33 	&	71600.07 	&	69933.31 	&		&	74269.66 	&	74269.75 	\\
51	&	71755.19 	&	71754.70 	&		&		&		&	74419.63 	\\
52	&	71912.89 	&		&		&		&		&		\\
53	&	72073.49 	&		&		&		&		&		\\
\hline
\hline
\end{tabular}
\end{scriptsize}
\end{center}
\end{table*}

\renewcommand\arraystretch{1}
\begin{table*}[htp]
\begin{center}
\begin{small}
\caption{Level energies (in vacuum \wn) of perturber states for \As\ $v=2,3,4$ states, with applied energy corrections obtained from the laser data.}
\label{perturbingstatelevelenergies}
\begin{tabular}
{r@{\hspace{10pt}}c@{\hspace{10pt}}c@{\hspace{15pt}}r@{\hspace{10pt}}c@{\hspace{10pt}}c@{\hspace{15pt}}r@{\hspace{10pt}}c@{\hspace{10pt}}c}
\hline
\hline\\[-2ex]
$J'$	&	\multicolumn{2}{l}{$d^3\Delta (v=7)$}		&	$J'$	&	\multicolumn{2}{l}{$e^3\Sigma^- (v=4)$}			&	$J'$	&		\multicolumn{2}{l}{$I^1\Sigma^- (v=3)$}		\\
42	&	F3f	&	70480.81	&	24	&	F1e	&	68633.43	&	6	&	F1f	&	67747.52	\\
43	&	F3f	&	70580.41	&	25	&	F1e	&	68693.40	&	7	&	F1f	&	67764.63	\\
39	&	F2e	&	70084.11	&	26	&	F1e	&	68748.20	&	8	&	F1f	&	67783.52	\\
39	&	F2f	&	70084.09	&	27	&	F2f	&	68879.09	&	9	&	F1f	&	67805.30	\\
	&		&			&	28	&	F2f	&	68948.57	&	&&			\\
&\multicolumn{2}{l}{$d^3\Delta (v=8)$}		&	29	&	F2f	&	69009.82	&	&\multicolumn{2}{l}{$I^1\Sigma^- (v=6)$}		\\
26	&	F3e	&	70160.00	&	30	&	F2f	&	69083.27	&	23	&	F1f	&	71297.30	\\
26	&	F3f	&	70160.01	&	32	&	F2f	&	69234.24	&	&&\\
	&		&			&	31	&	F3e	&	69235.70	&	&\multicolumn{2}{l}{$a'^3\Sigma^+ (v=13)$}	\\
&\multicolumn{2}{l}{$d^3\Delta (v=10)$}		&	32	&	F3e	&	69307.70	&	30	&	F2e	&	70525.76	\\
45	&	F2f	&	73560.81	&	33	&	F3e	&	69392.57	&	31	&	F2e	&	70591.85	\\
	&		&			&		&		&			&	33	&	F3f	&	70813.76	\\
&\multicolumn{2}{l}{$D^1\Delta (v=3)$}		&&\multicolumn{2}{l}{$e^3\Sigma^- (v=7)$}		&	34	&	F3f	&	70889.65	\\
47	&	F1e	&	71154.81	&	34	&	F1e	&	72248.71	&		&		&		\\
	&		&			&	36	&	F2f	&	72491.03	&	&\multicolumn{2}{l}{$a'^3\Sigma^+ (v=15)$}		\\
	&		&			&	37	&	F2f	&	72570.80	&	46	&	F2e	&	73692.42		\\
	&		&			&	40	&	F3e	&	72932.71	&		&		&		\\
\hline
\hline
\end{tabular}
\end{small}
\end{center}
\end{table*}

\section{Perturbation analysis}
\label{perturbation}

\begin{figure}
\begin{center}
\resizebox{1\textwidth}{!}{\includegraphics{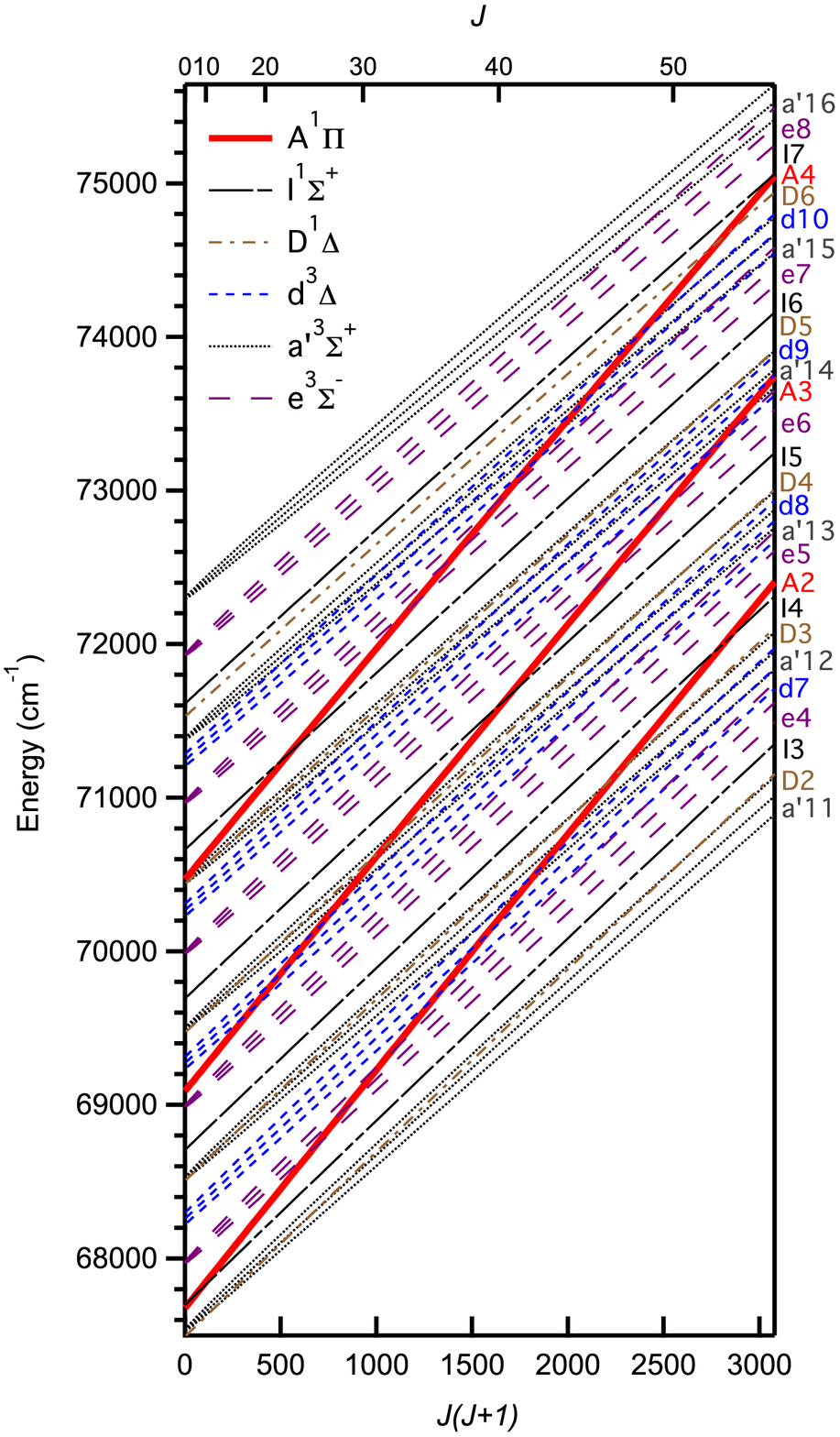}}
\caption{Perturbation diagram. Level energies as function of $J(J+1)$ for \As\ and perturber states. The labels denote the electronic state and vibrational quantum number, e.g. $a'16$ is shorthand for a$'^3\Sigma^+$ $(v=16)$.}
\label{fig:level}
\end{center}
\end{figure}

The CO \AX\ system is heavily perturbed by many other electronically excited states.
The \As\ (\2) levels are perturbed by levels of the \es\ (\4), \ds\ (\7), \as\ $(v=11,12)$, \Ds\ (\3) and \Is\ $(v=3,4)$ states;
the \As\ (\3) levels are perturbed by \es\ $(v=5,6)$, \ds\ (\8), \as\ $(v=13)$, and \Is\ (\5) states;
the \As\ (\4) levels are perturbed by \es\ (\7), \ds\ $(v=9,10)$, \as\ $(v=14,15)$, and \Is\ (\6) states.
Figure~\ref{fig:level} plots level energies as function of $J(J+1)$ for vibrational progressions of the \As\ and perturber states relevant to this study, showing the crossing points where local perturbations may occur.
The labels denote the electronic state and vibrational quantum number.
As will be discussed below, observable effects from some perturber states which do not cross with the \As\ state are manifest in the analysis, e.g. \as\ $(v=16)$.

We model each observed \AX\ band and interacting levels with a local deperturbation analysis, in a similar style to what was done by Niu. \emph{et. al.}~\cite{Niu2013}.
In this study, we use accurate transition energies obtained from FT spectroscopy for levels up to $J''=52$ of the \AX\ (2,0), (3,0) and (4,0) bands, as well as transitions attributed to perturber states.
The more accurate low-$J$ transition frequencies from the laser-based experiments~\cite{Niu2015} are preferentially used.
In order to perform a more comprehensive perturbation analysis, we also use results from previous investigations.
For the perturber state transitions, we used the low-$J$ data from Ref.~\cite{Herzberg1970} for \dX\ (7,0), (8,0) and (10,0); for \eX\ (4,0) and (5,0) data from Ref.~\cite{Morton1994}; for \eX\ (6,0) and (7,0) data from Ref.~\cite{Simmons1971}.
Since these older spectroscopic investigations are less accurate, we used relative weights that reflect the respective accuracies.
A relative weight of 10 is assigned to data from Niu~\cite{Niu2015}, 1 to the present FT data, 0.5-0.25 to weak and blended lines in the present FT data, 0.2 to data from Morton and Noreau~\cite{Morton1994}, and 0.1 to data from Herzberg \emph{et al.}~\cite{Herzberg1970} and Simmons and Tilford~\cite{Simmons1971}.

We performed a perturbation analysis on CO \AX\ $(2,0)$, $(3,0)$ and $(4,0)$ bands using the Pgopher software~\cite{pgopher}, where the same effective Hamiltonian model as Ref.~\cite{Niu2013} was used, retaining their symbols for the various molecular constants (see Table 6 of Ref.~\cite{Niu2013}).
The unweighted residuals of the fits are dominated by the uncertainties in the literature data.
The mean weighted residuals of the fits on the transition energies of the \AX\ $(2,0)$, $(3,0)$ and $(4,0)$ bands, including relevant literature data are 0.016~\wn, 0.014~\wn\ and 0.021~\wn, respectively.
In total, 235, 218, and 213 transitions are used to fit 22, 17, and 20 molecular fit parameters for the \As\ $(v=2,3,$ and $4)$ states, respectively.
The interaction parameters between \As\ states and perturbing states are denoted by $\eta_i$ for triplet perturbers and $\xi_i$ for singlet perturbers, where the $i=2,3,4$ indices correspond to the \As\ $(v=2, v=3$ and $v=4)$ levels.

In addition to the \AX\ $(2,0)$, $(3,0)$ and $(4,0)$ bands, we also improve the molecular constants for those perturber states for which a sufficient number of transitions are observed in the present FT experiment.
These extra lines are listed in Table~\ref{perturbingstatetransition} and, in general, occur at the level crossings of \As\ $(v)$ rotational series with those of perturber states.

The final set of deperturbed molecular constants obtained from the fits are summarized in Table~\ref{Molecule constants}. Molecular constants with an uncertainty indicated in parentheses () are fitted (free) parameters. The others are taken from previous deperturbation models, indicated accordingly in the footnote, and used as fixed parameters during the fitting procedures.
Note that we include all possible perturber states, even those that have no crossings with \As\ states (e.g. d$^3\Delta$ $(v=9)$ and a$'^3\Sigma^+$ $(v=11)$ states), to have a consistent model that facilitates comparisons with previous investigations, such as Ref.~\cite{Field-thesis}.

As expected from the extensive coverage of transition energies in the \AX\ bands, we obtain accurate molecular constants for the \As\ states.
This is also the case for the molecular constants of the \es$(v=4,7)$, \ds$(v=7)$, \as$(v=13)$, and \Is$(v=3)$ states as would be expected by inspection of Table~\ref{perturbingstatetransition}.

The band origin $T_v$ for \As\ $(v=3)$ obtained in the analysis has a larger uncertainty compared to $v=2$ and $4$, which is attributed to uncertainty regarding its perturbation by the \es\ $(v=5)$ state. The  rotational progression of \es\ $(v=5)$ does not actually cross with \As\ $(v=3)$ and no extra perturber transitions are observed.
Hence, less accurate literature values for transition energies in the \eX\ $(5,0)$ band~\cite{Morton1994} were included. The effect of this interaction is a global energy shift of \As\ $(v=3)$, which ultimately translates to a larger uncertainty of $T_v$ in \As\ $(v=3)$.

From Fig.~\ref{fig:level}, the D$^1\Delta$ $(v=4, 5, 6)$ states are expected to cross the A$^1\Pi$ vibrational states, however, no extra lines are obtained in this work, due to the small interaction parameters involved. Thus the inclusion of these states does not lead to any significant improvements in the quality of the fit. The addition of the I$^1\Sigma^-$ $(v=7)$ state, where the crossing is predicted to occur at $J=55$ and outside the data range, also does not improve the fitting. Therefore these states are excluded in the final model reported.

The \As\ $(v=4)$ state is perturbed by \ds\ $(v=10)$ at $J=45$ and \as\ $(v=15)$ at $J=46$, but these perturber lines are assigned to observed weak features and their assignment is tentative. The assigned transition P(46), Q(45) and R(44) in \AX\ (4,0) band are off in the resulting fit by 0.5 \wn. Hence, in the final fitting, the relative weights for these lines was set at 0.01.

The \ds\ $(v=9)$ and \as\ $(v=11)$ states perturb the \As\ $(v=4)$ and $(v=2)$ states, but without crossing them.
The interaction parameters are then strongly correlated with the $T_v$ parameters for the \As\ states. Hence, the interaction parameter for these two perturbations were fixed using the values from the original analysis of Field~\cite{Field-thesis}.

\section{Discussion and Conclusion}

High-precision frequency measurements of more than 450 rotational lines in CO \AX\ $(2,0)$, $(3,0)$ and $(4,0)$ bands (for $J$ up to 52) have been performed.
Three different configurations are used in the experiment to obtain the accurate transition frequencies for levels extending to high$-J$.
The accuracy of absolute transition frequencies is 0.01-0.02 \wn.
The present data, including recent laser data~\cite{Niu2015} as well as literature values are used to perform a successful global analysis of the perturbations by other electronic states.

In comparison to the original perturbation analysis by Field~\cite{Field-thesis}, the present investigation finds more local perturbation crossings, like those involving \es\ $(v=6)$, \ds\ $(v=10)$, \as\ $(v=12, 15)$, and \Is\ $(v=4, 5)$. These crossings are found at high rotational quantum number, which could be observed in our high temperature and saturated spectrum.
Molecular constants $T_v$ represent the deperturbed level energy separations between ground state \Xs\ $(v=0, J=0)$ and excited state $(v, J=0)$, and can be compared with the deperturbed $G(v)$ of Field~\cite{Field-thesis, Field1972a}. This yields good agreement at the 0.1 \wn\ level. 
Note that $T_v$ should not be compared directly with $E(v)$ in Field~\cite{Field-thesis, Field1972a}, since $E(v)$ is defined as the deperturbed level energy of the lowest existing rotational level in the particular excited state, which for the \As\ state is $J=1$.
Overall the extended data set and the improved accuracy of the level energies yields the derivation of molecular constants at a higher accuracy than in the previous analysis.
Values for the $B$ constants are found to be the same within $\sim10^{-4}$ \wn.
The interaction parameters $\eta$ are similar to those derived previously~\cite{Field-thesis, Field1972a}; here a different definition of interaction matrix elements should be considered, with the present numbers divided by a factor $\sqrt{3}$~\cite{pgopher}.

The accurate transition frequencies in A$^1\Pi$ and perturber states will be useful in the analysis of the astronomical spectra in order to determine a value for the cosmic microwave background temperature at high redshift~\cite{Noterdaeme2010}.
The results presented here are also relevant to studies that probe for a possible variation of the proton-to-electron mass ratio ($\mu$) using CO~\cite{Salumbides2012}.
Work is in progress in the analysis of CO A$^1\Pi-$X$^1\Sigma^+$ spectra toward the quasar J1237+064 combined with H$_2$ analysis in the same absorption system at redshift $z=2.69$~\cite{Dapra2015}.
This work provides more accurate laboratory wavelengths for the comparisons, and in addition the present perturbation analysis will be useful in improving the calculation of sensitivity coefficients for $\mu$-variation.

\smallskip

This work was supported by Dutch Astrochemistry Program of NWO (CW-EW).
We are grateful to the general and technical staff of SOLEIL for providing beam
time under project n$^{\rm{o}}20120653$.

\renewcommand\arraystretch{1}
\LTcapwidth=1\textwidth
\begin{scriptsize}
\begin{ThreePartTable}
\begin{TableNotes}
\item[a] Data from Ref.~\cite{Field-thesis} and converted.
\item[b] Constant fixed to that of \es, \3~\cite{Lefloch1987} as first-order approximation.
\item[c] Constant fixed to that of \ds, \5~\cite{Lefloch1987} as first-order approximation.
\item[d] Data from Ref.~\cite{Simmons1971}.
\item[e] Constant fixed to that of \as, $v=11$ as first-order approximation.
\item[f] Constant fixed to that of \as, $v=9$~\cite{Lefloch1987} as first-order approximation.
\item[g] Extrapolated from \Ds, \1\ and \2~\cite{Lefloch1987}.
\item[h] Constant fixed to that of \Ds, \1~\cite{Lefloch1987} as first-order approximation.
\item[i] Constant fixed to that of \Is, \2~\cite{Lefloch1987} as first-order approximation.
\end{TableNotes}
\begin{longtable}[htp]{l@{\hspace{15pt}}r@{\hspace{15pt}}r@{\hspace{15pt}}r@{\hspace{15pt}}r@{\hspace{15pt}}r}
\caption{Compilation of the molecular constants for the \As, \2, \3, and \4\ states of \CO\ and all perturbing states following from the present analysis. All values in vacuum \wn. 1$\sigma$ uncertainties given in parentheses in units of the least significant digit.}
\label{Molecule constants}
\\
\toprule\\[-2.5mm]
Singlet states	&\multicolumn{1}{c}{\As(\2)} &\multicolumn{1}{c}{\As(\3)}	&\multicolumn{1}{c}{\As(\4)}  &&\\[0mm]
\hline\\[-2.5mm]
\endfirsthead

\hline
\multicolumn{4}{c}{{continued on next page}} \\
\endfoot

\hline
\insertTableNotes
\endlastfoot

$T_v$	&	67675.9408	(6)	&	69088.368	(18)	&	70465.956	(7)	&			&			\\
$B$	&	1.55829	(1)	&	1.53503	(4)	&	1.51171	(2)	&			&			\\
$q$ $\times 10^6$	&	-7	(4)	&	-4	(3)	&	-10	(4)	&			&			\\
$D$ $\times 10^6$	&	7.55	(1)	&	7.68	(3)	&	7.91	(2)	&			&			\\
$H$ $\times 10^{11}$	&	-2.7	(3)	&	-2.2	(8)	&	-0.3	(6)	&			&			\\[2mm]
\hline\\[-2.5mm]
Triplet states	&\multicolumn{1}{c}{\es(\4)} &\multicolumn{1}{c}{\es(\5)} &\multicolumn{1}{c}{\es(\6)}	&\multicolumn{1}{c}{\es(\7)}	&\\[0mm]
\hline\\[-2.5mm]	
$T_v$	&	67969.82	(2)	&	68987.42	(2)	&	69986.25	(4)	&	70965.16	(2)	&			\\
$B$	&	1.20441	(5)	&	1.1877	\tnote{a}	&	1.16990	(6)	&	1.15289	(6)	&			\\
$\lambda$	&	0.69	\tnote{a}	&	0.63	\tnote{a}	&	0.64	\tnote{a}	&	0.76	\tnote{a}	&			\\
$D$ $\times 10^6$	&	6.35	(4)	&	6.664	\tnote{a}	&	6.637	\tnote{a}	&	6.22	(4)	&			\\
$H$ $\times 10^{12}$	&	-2	\tnote{b}	&	-2	\tnote{b}	&	-2	\tnote{b}	&	-2	\tnote{b}	&			\\
$\eta_2$	&	12.93	(1)	&			&			&			&			\\
$\eta_3$	&			&	9.4	(3)	&	11.37	(5)	&			&			\\
$\eta_4$	&			&			&			&	7.36	(3)	&			\\[2mm]
\hline\\[-2.5mm]
Triplet states	&\multicolumn{1}{c}{\ds(\7)} &\multicolumn{1}{c}{\ds(\8)}  &\multicolumn{1}{c}{\ds($v=9$)}	&\multicolumn{1}{c}{\ds($v=10$)}	&\\[0mm]
\hline\\[-2.5mm]																	
$T_v$	&	68257.71	(4)	&	69270.92	(1)	&	70266.034	\tnote{a}	&	71242.54	(6)	&			\\
$B$	&	1.18277	(9)	&	1.16629	(3)	&	1.15010	\tnote{a}	&	1.13312	(3)	&			\\
$A$	&	-16.82	(3)	&	-17.16	(2)	&	-17.34	\tnote{a}	&	-17.28	(4)	&			\\
$\lambda$	&	1.15	\tnote{a}	&	1.2	\tnote{a}	&	1.31	\tnote{a}	&	1.58	\tnote{a}	&			\\
$\gamma$ $\times 10^3$	&	-9	\tnote{a}	&	-8	\tnote{a}	&	-10	\tnote{a}	&	-8	\tnote{a}	&			\\
$D$ $\times 10^6$	&	6.44 	(4)	&	6.41	\tnote{a}	&	6.40	\tnote{a}	&	6.53	(2)	&			\\
$H$ $\times 10^{12}$	&	-0.8	\tnote{c}	&	-0.8	\tnote{c}	&	-0.8	\tnote{c}	&	-0.8	\tnote{c}	&			\\
$A_D$ $\times 10^4$	&	-1	\tnote{a}	&	-1	\tnote{a}	&	-1	\tnote{a}	&	-1	\tnote{a}	&			\\
$\eta_2$	&	10.74	(2)	&			&			&			&			\\
$\eta_3$	&			&	0.79	(7)	&			&			&			\\
$\eta_4$	&			&			&	7.00	\tnote{a}	&	-1.6	(2)	&			\\[2mm]
\hline\\[-2.5mm]
Triplet states	&\multicolumn{1}{c}{\as($v=11$)}	&\multicolumn{1}{c}{\as($v=12$)}	&\multicolumn{1}{c}{\as($v=13$)}	&\multicolumn{1}{c}{\as($v=14$)}	&\multicolumn{1}{c}{\as($v=15$)}\\[0mm]
\hline\\[-2.5mm]																				
$T_v$	&	67529.52	\tnote{d}       	&	68519.7	\tnote{d}       	&	69491.43	(6)	&	70443.55	\tnote{d}       	&	71377.83	(2)	\\
$B$	&	1.14921	\tnote{a}	&	1.1338	\tnote{d}       	&	1.11772	(5)	&	1.1051	\tnote{a}	&	1.08596	\tnote{a}	\\
$\lambda$	&	-1.103	\tnote{a}	&	-1.103	\tnote{e}	&	-1.151	\tnote{a}	&	-1.14	\tnote{a}	&	-1.07	\tnote{a}	\\
$\gamma$ $\times 10^3$	&	4.47	\tnote{a}	&	0	\tnote{a}	&	0	\tnote{a}	&	0	\tnote{a}	&	0	\tnote{a}	\\
$D$ $\times 10^6$	&	6.255	\tnote{a}	&	6.251	\tnote{a}	&	6.254	\tnote{a}	&	6.263	\tnote{a}	&	6.284	\tnote{a}	\\
$H$ $\times 10^{12}$	&	-0.4	\tnote{f}	&	-0.4	\tnote{f}	&	-0.4	\tnote{f}	&	-0.4	\tnote{f}	&	-0.4	\tnote{f}	\\
$\eta_2$	&	6.81	\tnote{a}	&	5.82	(3)	&			&			&			\\
$\eta_3$	&			&			&	7.06	(2)	&			&			\\
$\eta_4$	&			&			&			&	8.24	(4)	&	-6.93	(5)	\\[2mm]
\hline\\[-2.5mm]
Singlet states	&\multicolumn{1}{c}{\Ds(\3)}	&&&&\\[0mm]
\hline\\[-2.5mm]																				
$T_v$	&	68504.34	(2)	&			&			&			&			\\
$B$	&	1.19	\tnote{g}	&			&			&			&			\\
$D$ $\times 10^6$	&	7	\tnote{h}	&			&			&			&			\\
$H$ $\times 10^{12}$	&	-0.3	\tnote{h}	&			&			&			&			\\
$\xi_2$	&	0.0331	(3)	&			&			&			&			\\[2mm]
\hline\\[-2.5mm]
Singlet states	&\multicolumn{1}{c}{\Is(\3)}	&\multicolumn{1}{c}{\Is(\4)}	&\multicolumn{1}{c}{\Is(\5)}	&\multicolumn{1}{c}{\Is(\6)}&	\\[0mm]
\hline\\[-2.5mm]																					
$T_v$	&	67696.79	(3)	&	68706.08	\tnote{d}	&	69692.39	\tnote{d}	&	70661.38	(3)	&			\\
$B$	&	1.2069	(4)	&	1.1915	\tnote{d}	&	1.1748	\tnote{d}	&	1.1568	\tnote{d}	&			\\
$D$ $\times 10^6$	&	6.89	\tnote{i}	&	6.89	\tnote{i}	&	6.89	\tnote{i}	&	6.89	\tnote{i}	&			\\
$H$ $\times 10^{12}$	&	3	\tnote{i}	&	3	\tnote{i}	&	3	\tnote{i}	&	3	\tnote{i}	&			\\
$\xi_2$	&	0.062	(1)	&	0.079	(2)	&			&			&			\\
$\xi_3$	&			&			&	0.0338	(6)	&			&			\\
$\xi_4$	&			&			&			&	0.0193	(7)	&			\\
\end{longtable}
\end{ThreePartTable}
\end{scriptsize}

\bibliographystyle{tMPH}
\bibliography{CompleteDataBase}

\end{document}